# OpenMS WebApps: Building User-Friendly Solutions for MS Analysis

Tom David Müller †[1,2], Arslan Siraj †[1,2], Axel Walter[1,2], Jihyung Kim[1,2], Samuel Wein[1,2], Johannes von Kleist[1], Ayesha Feroz[1,2], Matteo Pilz[1,2], Kyowon Jeong[1,2], Justin Cyril Sing[3,4], Joshua Charkow[3,4], Hannes Luc Röst*[3,4], Timo Sachsenberg*[1,2]

† Equal Contribution

* corresponding authors: hannes.rost@utoronto.ca, timo.sachsenberg@uni-tuebingen.de

1. Applied Bioinformatics, Department of Computer Science, University of Tübingen, Tübingen, Germany
2. Institute for Bioinformatics and Medical Informatics, University of Tübingen, Tübingen, Germany
3. Donnelly Centre for Cellular and Biomolecular Research, University of Toronto, Toronto, Ontario M5S 3E1, Canada
4. Department of Molecular Genetics, University of Toronto, Toronto, Ontario M5G 1A8, Canada

## ABSTRACT

Liquid Chromatography Mass Spectrometry (LC-MS) is an indispensable analytical technique in proteomics, metabolomics, and other life sciences. While OpenMS provides advanced open-source software for MS data analysis, its complexity can be challenging for non-experts. To address this, we have developed OpenMS WebApps, a framework for creating user-friendly MS web applications based on the Streamlit Python package. OpenMS WebApps simplifies MS data analysis through an intuitive graphical user interface, interactive result visualizations, and support for both local and online execution. Key features include workspaces management, automatic generation of input widgets, and parallel execution of tools resulting in high performance and ready-to-use solutions for online and local deployment. This framework benefits both researchers and developers: scientists can focus on their research without the burden of complex software setups, and developers can rapidly create and distribute custom WebApps with novel algorithms. Several applications built on the OpenMS WebApps template demonstrate its utility across diverse MS-related fields, enhancing the OpenMS eco-system for developers and a wider range of users. Furthermore, it integrates seamlessly with third-party software, extending benefits to developers beyond the OpenMS community.

## INTRODUCTION

Liquid Chromatography Mass Spectrometry (LC-MS) has become an indispensable analytical technique for the comprehensive analysis of complex biological samples in proteomics, metabolomics and other life science disciplines. The analysis and interpretation of MS data requires advanced software solutions which can be non-trivial even for experts in the field. OpenMS, an open-source software framework, offers a wide range of algorithms, through a C++ library, Python bindings (pyOpenMS) and command line (TOPP) tools for MS data processing [1], [2]. Integration into modular workflow systems allows for a high degree of control over the analysis process, making it ideal for complex workflows. This flexibility comes with a cost: non-expert users may find it challenging to use this modularity for routine laboratory tasks. Applications built on web technologies have seen a steep rise in popularity in recent years. In contrast to classic desktop GUI applications, web applications can be used online or on the intranet without installation or dependence on the user's operating system. Some types of web applications also support local execution on the user's computer. The latter can be relevant if there are restrictions regarding the upload of private data into the cloud or limited network bandwidth. Other LC-MS tools have successfully adopted web applications to enhance accessibility, such as the GNPS Dashboard [3], MassDash [4], µSpotReader [5], and the FBMN Stats GUIde [6]. Such user-friendly web-based interfaces make complex data analysis tasks more accessible to a broader user base.

OpenMS WebApps build upon this trend, offering a framework optimized for creating user-friendly MS analysis applications based on OpenMS and other third-party tools. In the following sections, we will describe the methods used to develop the OpenMS WebApps template, discuss its key features and benefits for users and developers, and present several example applications built with it.

## METHODS

Given the plethora of web frameworks available, no one-size-fits-all solution exists. Based on several empirical experiments with different Python-based frameworks, we struck a balance and selected one that is powerful enough to address common use cases in MS-based research software development and at the same time works well for a wide range of developers with diverse backgrounds. Streamlit is an open-source framework popular in bioinformatics web applications [7], which suits students with limited exposure to Python and web programming as well as experienced and more seasoned programmers. It comes with predefined user interface elements such as multi-page app support, a sidebar, and input widgets. It is actively developed and integrates well with key Python libraries such as Pandas and Plotly [8] for interactive tables and figures.

OpenMS WebApps is based on Streamlit 1.38 and offers an opinionated template with a unified file and folder structure that follows the key steps of typical workflows. Its architecture is shown in Fig. 1. Basic security features like captcha control, a start page with a quick start guide, global settings, data upload with workspaces to organize project-specific files, configuration of workflow parameters, parallel execution of tools or scripts, continuous logging, and results visualization (see Fig. 2 and 4) are made available as part of a web app template repository on GitHub. Several pages, each corresponding to one of the key steps, provide a clear structure for users. They are accompanied by documentation for developers.

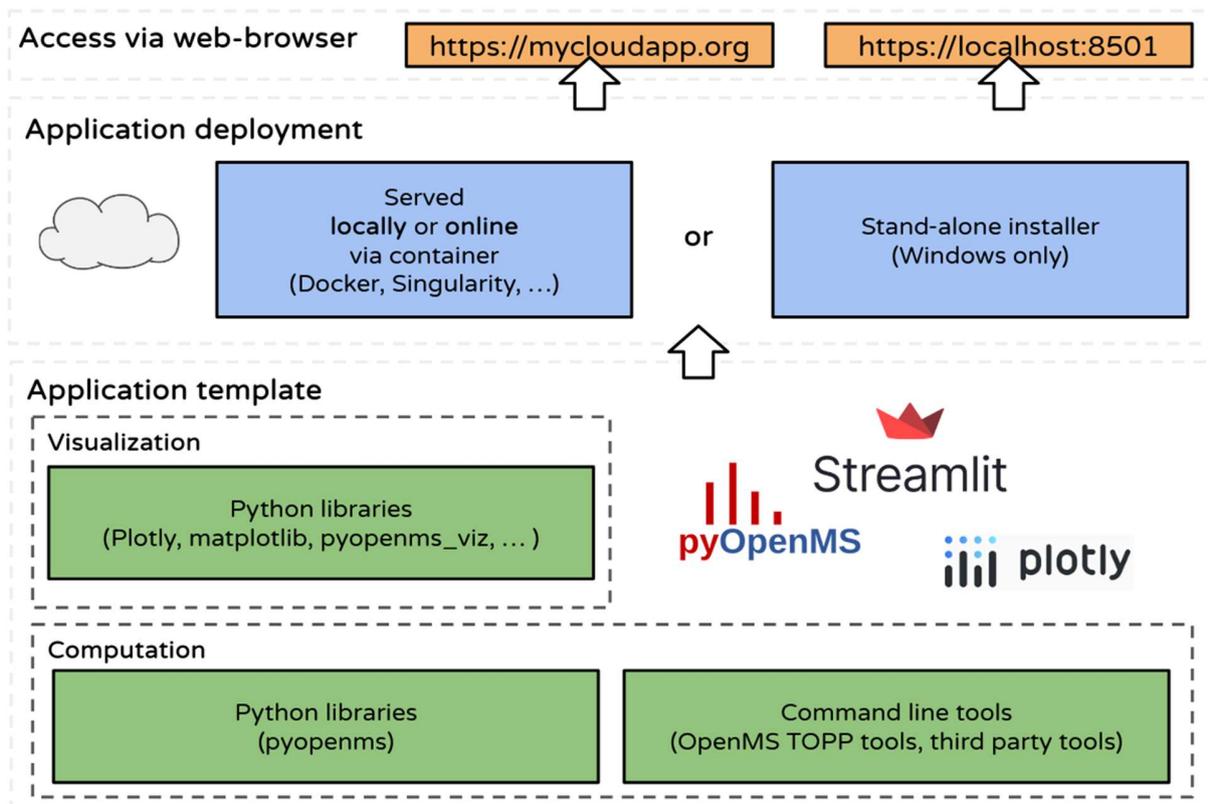

**Figure 1. Overview:** The app template is based on the Streamlit library. Computation of mass spectrometry analysis workflows is done either via Python libraries or via command line tools such as the OpenMS and other third-party tools. The app can be served via containers in the cloud or on a local network. Apps can be run on a local Windows PC via executables generated by a GitHub action including all necessary dependencies in a stand-alone installer.

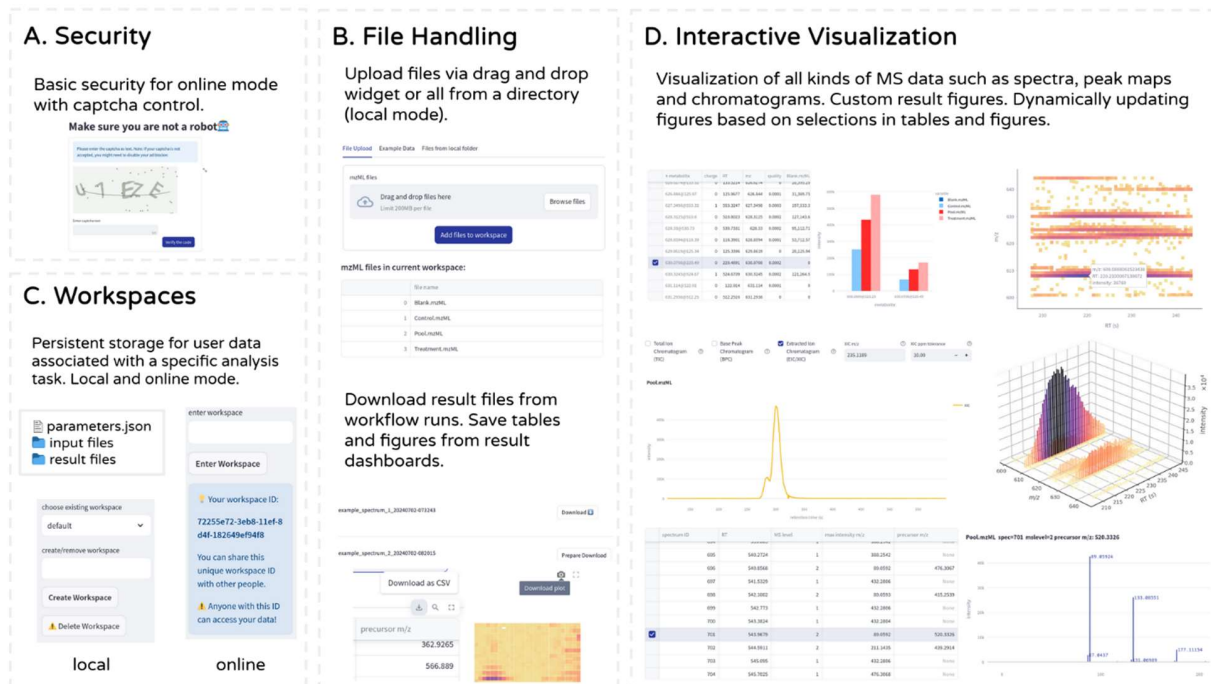

**Figure 2. Features. A: Security:** Captcha control prevents unwanted access to the app. **B: File Handling:** Files can be uploaded to the workspace via an upload widget or by specifying a local directory path. Result files can be downloaded as zip files. Furthermore, tables and figures from the interactive result dashboards can be downloaded directly. **C. Workspaces:** Workspaces manage uploaded and analysis result files. **D. Interactive Visualization:** MS data visualizations can be dynamically updated based on user selections.

One requirement we aimed to fulfill was support for two deployment modes: (1) as a web-based application running on a server and (2) as a standalone application executable on a local PC. Some of the features provided by the template have different behavior depending on online or local execution. For instance, a captcha control widget is only displayed if users access the application online. The file upload step allows to add files to the project workspace. If executed online, files are uploaded and stored in the workspace folder. For local execution, the files are copied to the workspace by default, however, to save on disk space system links can be used instead. Other parts, e.g., the configuration step that typically follows file upload, offers a user interface to configure the workflow, tool or script parameters using visual components such as textboxes, checkboxes, drop-down lists, and sliders. OpenMS tools are particularly well supported with automatic parameter handling and input widget generation. Third-party command line tools or Python scripts can conveniently be integrated but require manual specification of their parameters. The execution step allows users to run the configured workflow by pressing a single button. During workflow execution, a stop button is shown that allows to abort the workflow. A log output widget displays diagnostic messages like command line output. To define the workflow in the first place, developers only need to implement four methods of the Workflow class. These methods directly translate to four pages, guiding the user through the workflow. On the first page, the user uploads input files for the workflow, the second page allows for the configuration of workflow parameters, and the third page provides a convenient interface to run the workflow. The final page is intended for visualizing results. Streamlit natively supports visualizations using Python plotting packages like Matplotlib, Plotly [8], or plotting packages specialized for MS data (e.g., spectrum_utils [9], [10] or pyopenms_viz [11]). In addition to these Python based visualizations, custom visualization components can be developed in web frameworks such as Vue.js and embedded within WebApps (see the FLASHViewer WebApp [12] as an example).

To ensure that all project-related files (input, configurations, results, state of the workflow) persist, even if the user closes the browser and revisits the web app later, we have expanded the Streamlit session management. By default, Streamlit stores session data only until a new session is assigned, which can lead to potential data loss. OpenMS WebApps addresses this by storing project and session data within project workspaces. When a workspace is created, the initial Streamlit session ID is strictly tied to the project and displayed to the user as a unique workspace ID. This ID is encoded in the URL and allows users to bookmark and recover their data and session state using the URL. To conserve resources on web servers, a script regularly deletes unused sessions by default. During local execution, users have full control over workspaces, including methods to create and delete them.

We assist developers with two preconfigured continuous integration (CI) pipelines for local and web deployment. The first GitHub action CI script enables developers to transform the Streamlit applications into Windows executables for local execution. Windows executables are packaged using embeddable Python 3.11 together with all required Python packages and command line tool binaries. Consequently, Windows users can run their app locally in the browser, enabling secure, accessible and easy control of their workspaces located on their PC. In the WebApps template, we added a button to download the installer for Windows directly on the web hosted app. This addresses the important use-case of researchers that want to try out an application online on their test data before they locally analyze e.g. data subject to access control. In this scenario, users download the stand-alone Windows application by clicking on the button in the online hosted web application after they reassured themself about the validity of processed test data.

To simplify deployment of one or multiple WebApps, we provide an example repository that illustrates how individual apps are organized in separate GitHub submodules and orchestrated utilizing Docker Compose for the online (or local network) deployment of one or multiple WebApps. Every application is assigned its own individual Docker volume for workspaces and is allocated unique ports on the host system forwarding to the application port in the container. For local deployment a GitHub action workflow creates packaged Windows executables (see Fig. 3 for details on the deployment process).

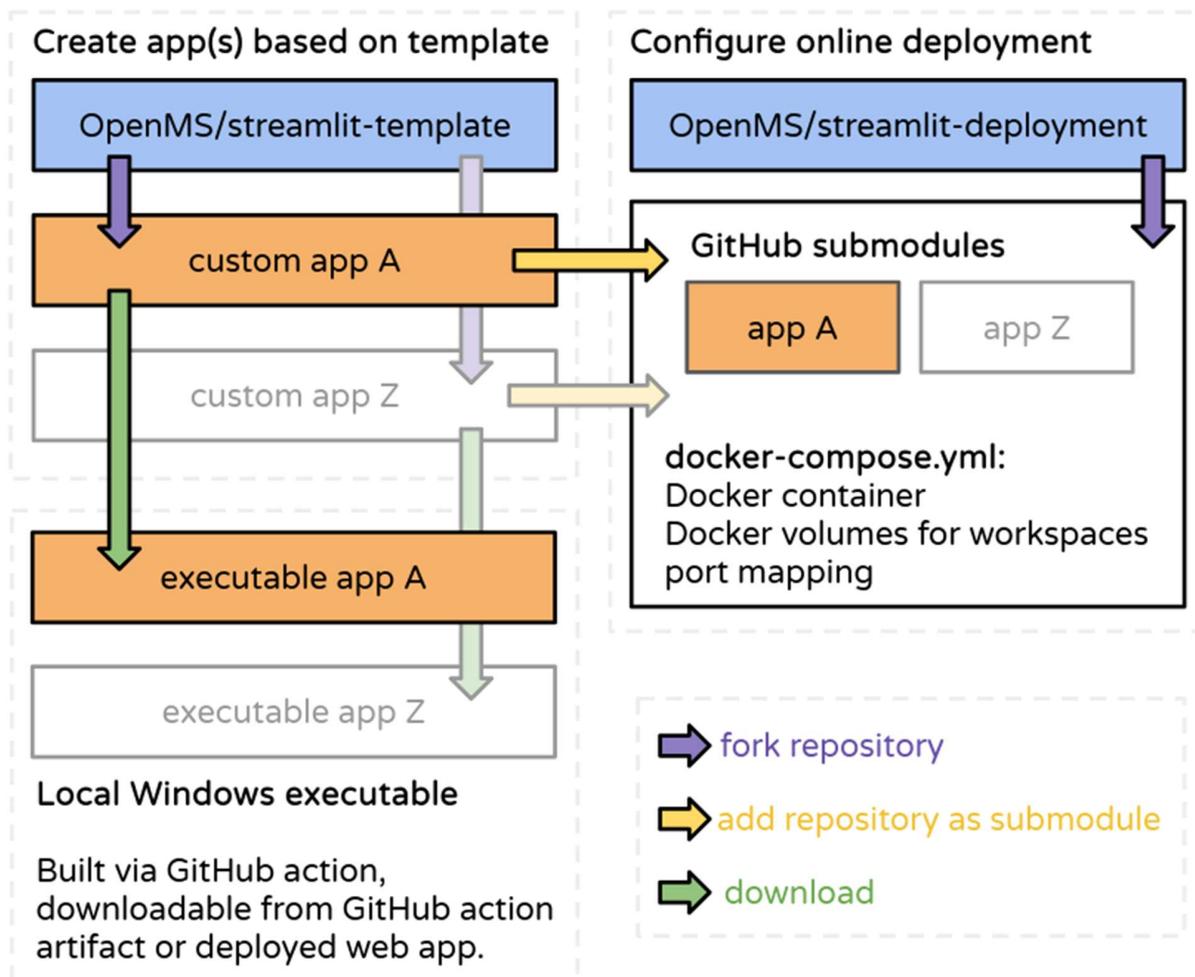

**Figure 3. Online and local deployment.** WebApps can be created by forking and customizing the OpenMS streamlit-template repository. A GitHub action workflow included in the template creates packaged Windows executables for local execution. For hosting one or multiple apps online or on local networks we provide the OpenMS streamlit-deployment repository. Adding WebApps as submodules to a fork of the deployment repository only requires adapting the docker-compose configuration file.

## RESULTS AND DISCUSSION

OpenMS offers a modular framework for building workflows for proteomics and metabolomics as well as emerging fields such as top-down proteomics, nucleic acid crosslinking and RNA-omics [2]. Currently available OpenMS workflows utilize workflow managers either based on scripting and command line tools (e.g. Nextflow [13], snakemake [14]) or visual workflow managers (e.g. KNIME [15], Galaxy [16]). While these platforms are powerful and versatile, they can create a significant barrier for non-experts and bench scientists, who possess invaluable domain knowledge but may find these tools challenging to use due to their complexity and technical demands. While building or customizing workflows in a graphical user interface remains an important use case, web applications may offer significant advantages for users and developers (see Tab. 1 for a summary), which will be discussed in the following sections.

**Table 1: OpenMS WebApps vs. other Workflow Systems. WebApps, Visual Workflows (e.g., KNIME [15], Galaxy [16]), and Command Line-Based Workflows (e.g., Nextflow [13], Snakemake [14]) from User and Developer Perspectives.**

|  | OpenMS WebApps | Visual workflows | Command line workflows |
|---|---|---|---|
| **User perspective** | | | |
| User experience | access via browser | simple online or desktop interface | typically console |
| Setup and installation | access via browser without installation or local Windows executable | access via browser or installation on multiple platforms | requires installation and configuration |
| Configuration | input through user-friendly forms | parameter settings via nodes or user-friendly forms | manual parameter settings, configuration files or assisted by other tools |
| Integrating user feedback during workflow execution (e.g., manual, visual filtering of intermediate results between workflow steps) | high degree; comparable to native desktop applications | typically requires workarounds or feels less seamless | typically not a use-case for high-throughput workflows |
| Visualization | highly customized, interactive | support for interactive visualizations vary between platforms | typically non-interactive |
| **Developer perspective** | | | |
| Programming | Python | no programming required for simple tasks; complex tasks or visualizations: typically | workflow language; custom tasks can be |

|  |  | Python, R, or java script nodes | implemented in scripts |
| --- | --- | --- | --- |
| Level of Customization | highest | lowest (standard nodes), high if scripting is included | high if scripting is included |
| Scalability | small to medium datasets | small to medium datasets | large datasets |
| Development Time | slower than visual workflows for simple tasks; faster for tasks that are complex or have specific visualization requirements | quick for simple tasks; complex workflows take longer. | Initial setup time-consuming; efficient once established. |

OpenMS WebApps are designed to offer, by default, an intuitive interface with a minimal learning curve, making the resulting WebApps particularly fitting for bench scientists who want to focus on biological questions rather than software technicalities. If hosted online, users can access apps instantly and from any device that can run modern web browsers. Users who only require web-based access can thus avoid the complexities of setup and local installation processes. Another significant benefit for users that exclusively access apps online, especially in larger institutions or enterprise environments, is their independence from specific hardware requirements (e.g., GPUs for AI tasks, high memory or CPU requirements) since computations are performed on dedicated hosts that can be tailored and optimized for specific applications. Additionally, in scenarios where data privacy is a concern hosting the applications in a private network and providing users only access to processed results but not the primary data reduces the risk of exposing sensitive information. One significant limitation, which paradoxically serves as a major advantage for user experience, is that web applications typically focus on a single type of analysis or workflow. This allows them to be designed with the user in mind, featuring intuitive input forms and tailored visualizations of results. A level of user experience that is challenging to achieve using general workflow platforms. A key concept of OpenMS WebApps is a simple workspace management system to organize project data and results. In online mode, workspaces can be shared with collaborators via a URL, to promote collaboration and accelerate research. In local mode, workspaces are organized in folders providing full control over the data to the user. Users can, by default, conveniently download figures by clicking a camera icon in the top right corner, while tabular results can be saved as CSV files via a download button. This simplifies sharing or further processing of results. In terms of scalability, OpenMS WebApps are currently most suitable for small to medium sized datasets which can be processed on a single computer. Long-running workflows are supported, allowing users to close the app and later return to an active or completed workflow run. However, the analysis of datasets that require distributing computation to multiple compute nodes is currently not supported.

Developers benefit from reduced development time to create user friendly web applications. If the execution sequence of tools or scripts of a workflow is known (including required parameters) a web application with only basic result visualization can typically be built in a day or two by a developer familiar with Python. This reduction in development time is accomplished by offering an app template with robust default settings and built-in methods. These methods allow developers to rapidly define file upload widgets, configuration options, workflow tools and steps, and result dashboards (see Fig. 4) using just a few lines of code.

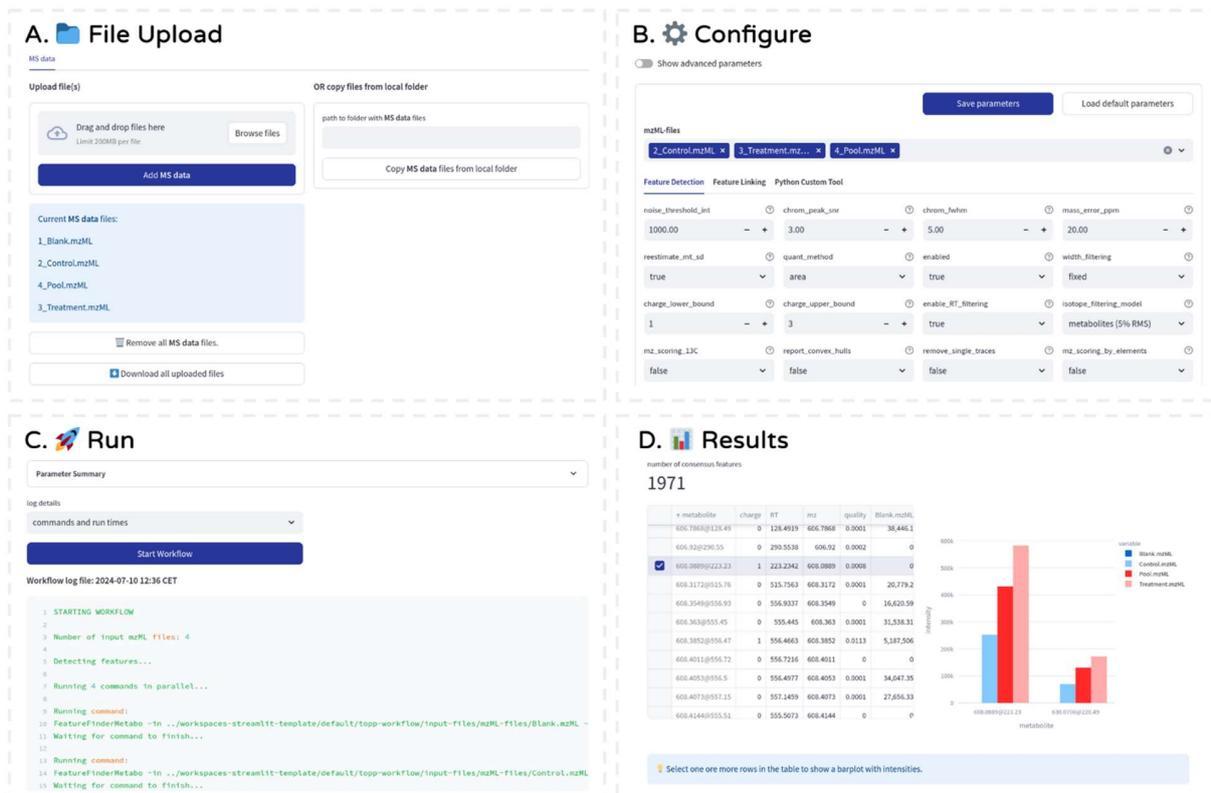

**Figure 4. A comprehensive framework for building performant WebApps with OpenMS TOPP tools, third party tools and Python scripts with an auto-generated user interface.** **A. File Upload:** Users can easily upload and organize their input files stored in the workflow directory. **B. Configure:** Users can set, adjust and save parameters for their workflows using automatically generated input widgets for TOPP tool-specific, Python script and custom parameters. **C. Run:** Continuously updated log with three detail levels for full insights. Users can exit the app and return to the workspace later while the workflow is running or already finished. **D. Results:** Interactive result dashboard from which tables, figures and result files can be downloaded.

The template implements a basic session or workspace management and provides the necessary boilerplate code and components developers can choose from. Developers can configure the app through a global settings file to enable optional components such as captcha control (see Fig. 2A, online mode only) or usage statistics tracking. The modular nature of the template app allows for easy modification, extension and customization. For instance, if required, the simple captcha control could be replaced with a more complex authentication solution. Continuous integration scripts are provided as part of the template, reducing the time required to implement application testing. The reduction in development time is most prominent if OpenMS tools are used because of the framework's native support for OpenMS parameter handling. In this case, the configuration page is built automatically. For non-OpenMS tools, developers benefit from all other parts of the template, but the configuration page requires manual creation of controls using a simple API (see Supplementary Fig. 1). The template further provides the necessary scripts to implement the two main deployment options: online/within local networks or execution on a single PC with Windows executables. Bioinformaticians in core facilities benefit from the ability to rapidly develop self-hosted WebApps tailored to specific projects. Tool or algorithm developers can utilize cloud-hosted WebApps to create demos with example datasets for new tools and algorithms (see Fig. 5). Because OpenMS WebApps can be released independently of the OpenMS release cycles (e.g., created from a developer branch) it results in zero delays for publishing.

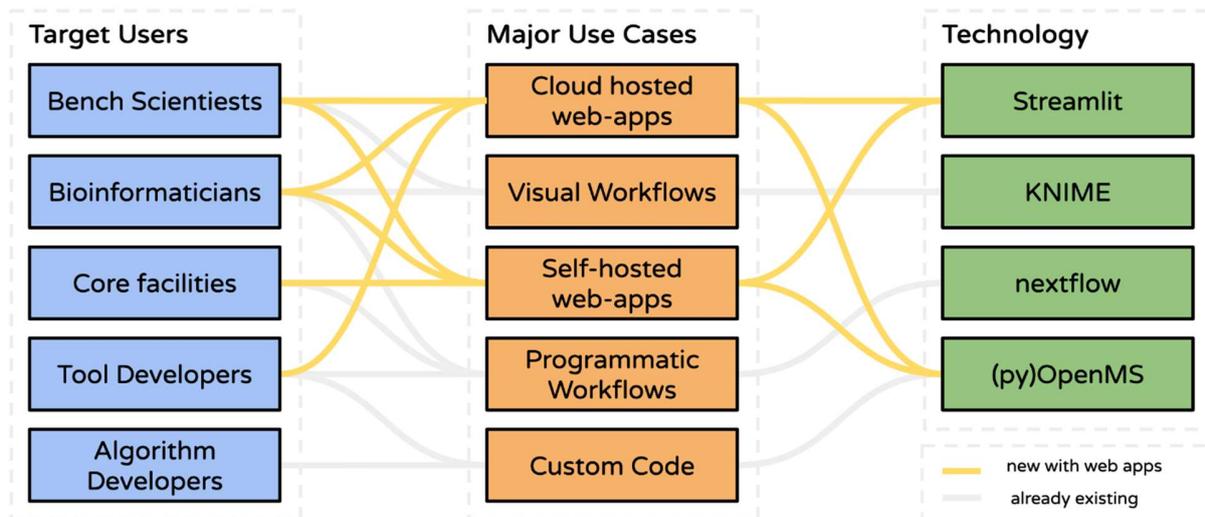

**Figure 5. Extending the OpenMS user base with WebApps:** Bench scientists can use WebApps to conduct MS analysis independently of assistance by bioinformaticians thanks to the accessible graphical user interface and interactive result dashboards. Bioinformaticians and core facilities can rapidly develop self-hosted WebApps tailored towards specific projects. Tool developers can utilize cloud-hosted WebApps as demos with example data sets for new tools and algorithms.

We created a variety of apps based on the template app to demonstrate how it seamlessly integrates with OpenMS tools. While some serve niche use cases, such as NuXL (nucleic acid crosslinking) [17] and NASEWEIS (NucleicAcidSearchEngine Web Execution In Streamlit) [18], [19], others address broader research fields like top-down proteomics (FLASHViewer) and metabolomics (UmetaFlow) (see Tab. 2 and supplementary information).

**Table 2: Currently available WebApps hosted by the OpenMS team.**

| **OpenMS template app** | |
|---|---|
| Description | A template app designed as a base for new OpenMS WebApps. Includes examples for building simple and complex workflows. Serves as documentation. |
| GitHub repository | https://github.com/OpenMS/streamlit-template |
| App URL | https://www.openms.org/webapps/streamlit-template/ |
| **NASEWEIS** | |
| Description | NASEWEIS allows users to upload data from an MS experiment, a FASTA file for (potentially modified) sequences to search and set parameters regarding the resolution of the instrument. It then identifies candidates for oligonucleotide spectral matches and does FDR to return a table of oligonucleotide hits. |
| GitHub repository | https://github.com/poshul/nase-weis |
| App URL | https://www.openms.org/webapps/naseweis/ |
| **UmetaFlow** | |
| Description | All in one metabolomics tool, UmetaFlow pipeline and extensive functionality for extracted ion chromatograms including mass calculator. Included raw data viewer and statistics. |
| Implementation | TOPP workflow framework for UmetaFlow pipeline; custom Python scripts for XIC and statistics; raw data viewer from template app |
| External Tools | SIRIUS [20], [21] |
| GitHub repository | https://github.com/axelwalter/umetaflow-gui |
| App URL | https://www.openms.org/webapps/umetaflow/ |
| **NuXL-app** | |
| Description | The NuXL-app is a web application version of the NuXL search engine, a specialized tool developed for the analysis of protein nucleic acid cross-linking mass spectrometry data. It allows for reliable, FDR-controlled assignment of protein–nucleic acid crosslinking sites from samples treated with UV light or chemical crosslinkers and offers user-friendly matched spectra visualization including ion annotations. |
| Implementation | TOPP workflow framework for NuXL; annotated spectra visualization |
| External Tools | Percolator [22] |
| GitHub repository | https://github.com/Arslan-Siraj/nuxl-app |
| App URL | https://www.openms.org/webapps/nuxl/ |
| **FLASHViewer** | |
| Description | A comprehensive viewer for Top-Down-Proteomics. All FLASH* tools can be run within the WebApp. Their output can be visualized using complex but highly comprehensive visualizations of identified proteoforms, sequence tags, fragment ions, post-translational-modifications, quantified spectra, deconvolved spectra, and raw spectra. |
| Implementation | TOPP workflows for running one or multiple TOPP tools in sequence. Raw outputs can be accessed directly by the user or processed automatically for visualization using custom Python scripts. An upload section for the visualization of externally run workflows is also provided. Visualizations were developed using Vue.js and embedded within the WebApp. Configuration of the layout of the visualizations is possible using dedicated configuration pages. |
| GitHub repository | https://github.com/t0mdavid-m/FLASHViewer |
| App URL | https://www.openms.org/webapps/flashviewer/ |
| **SagePTMScanner** | |
| Description | A web application version of the SageAdapter tool from OpenMS packaged with Sage, a tool developed for fast proteomics database searching. The WebApp offers many features such as FDR-control, retention time prediction and open searches. Output can be visualized both as spectra and discovered PTMs as a table and a graph. |
| Implementation | TOPP workflow framework for the Sage search engine, visualization of PTMs and spectra. |
| External Tools | Sage [23] |
| GitHub repository | https://github.com/JohannesvKL/SageAdapterApp |
| App URL | https://www.openms.org/webapps/SagePTMScanner/ |

We envision that WebApps for OpenMS-based MS workflows further extend the OpenMS user base, serve as working examples to research software developers working in computational MS, and facilitates the use of the template.

## CONCLUSION

OpenMS WebApps provides an accessible framework for rapid development of intuitive web applications for MS data analysis. Bench scientists without bioinformatics expertise benefit from an improved user experience and accessibility compared to the more flexible but also more complex workflow managers. Several apps have been developed, receiving overall positive user feedback. Bioinformaticians are empowered to rapidly develop and distribute custom WebApps, e.g. to provide novel tools and algorithms along with example datasets to their group or a wider audience. In general, we anticipate that these technologies have the potential to significantly lower the barrier to make novel tools available and to conduct computational MS analysis by a broader audience. This is particularly relevant in areas where no native desktop applications exist or where development costs would outbalance the benefits. While native desktop GUI applications can offer superior performance and tighter integration with the operating system, their development typically requires considerably more effort and specialized skills. The framework, in its current form, is based on opinionated decisions to provide a user- and developer-friendly solution for common use cases. As an open-source project, we actively encourage community contributions to help further improve the framework. By adding new features and extensions, contributors can help meet the evolving needs of users and developers and expand the framework beyond the OpenMS ecosystem.

## AUTHOR CONTRIBUTIONS

A.S., A.W., J.K., T.S. and T.M. conceived the work and implemented the software. All authors contributed to the writing of the manuscript. All authors read and approved the final manuscript.

## CODE AVAILABILITY

OpenMS WebApps code can be found in two GitHub repositories. One for the template application (https://github.com/OpenMS/streamlit-template) and one for the deployment (https://github.com/OpenMS/streamlit-deployment).

## ACKNOWLEDGMENTS

A.F. and A.S. are part of the MSCA-ITN-2020 PROTrEIN project, which received funding from the European Union's Horizon 2020 research and innovation program under the Marie Skłodowska-Curie grant agreement No: 956148. A.W. T.S. and T.M. acknowledge funding by the Federal Ministry of Education and Research in the frame of de.NBI/ELIXIR-DE (W-de.NBI-022). K.J. T.S. and T.M. are supported by the Ministry of Science, Research and Arts Baden-Württemberg. J.C.S. was supported by the European Research Area Network Personalized Medicine Cofund (PerProGlio, #506078) and CIHR (#507496). J.C. was supported by CIHR (#506078) and the Ontario Graduate Scholarship.

## COMPETING INTERESTS

T.S. and S.W. are officers in OpenMS Inc., a non-profit foundation that manages the international coordination of OpenMS development. All remaining authors declare no competing interests.

## REFERENCES

[1] H. L. Röst, U. Schmitt, R. Aebersold, and L. Malmström, "pyOpenMS: A Python-based interface to the OpenMS mass-spectrometry algorithm library," *PROTEOMICS*, vol. 14, no. 1, pp. 74–77, Jan. 2014, doi: 10.1002/pmic.201300246.

[2] J. Pfeuffer *et al.*, "OpenMS 3 enables reproducible analysis of large-scale mass spectrometry data," *Nat. Methods*, Feb. 2024, doi: 10.1038/s41592-024-02197-7.

[3] D. Petras *et al.*, "GNPS Dashboard: collaborative exploration of mass spectrometry data in the web browser," *Nat. Methods*, vol. 19, no. 2, pp. 134–136, Feb. 2022, doi: 10.1038/s41592-021-01339-5.

[4] J. C. Sing, J. Charkow, M. AlHigaylan, I. Horecka, L. Xu, and H. L. Röst, "MassDash: A Web-based Dashboard for Data-Independent Acquisition Mass Spectrometry Visualization," Jan. 17, 2024. doi: 10.1101/2024.01.15.575772.

[5] Functional-Metabolomics-Lab, *MicrospotReader: WebApp for the Detection of Bioactive Features in an Untargeted Metabolomics Experiment*. [Online]. Available: https://github.com/Functional-Metabolomics-Lab/MicrospotReader

[6] A. K. Pakkir Shah *et al.*, "The Hitchhiker's Guide to Statistical Analysis of Feature-based Molecular Networks from Non-Targeted Metabolomics Data," Nov. 01, 2023. doi: 10.26434/chemrxiv-2023-wwbt0.

[7] C. Nantasenamat, A. Biswas, J. M. Nápoles-Duarte, M. I. Parker, and R. L. Dunbrack, "Building bioinformatics web applications with Streamlit," in *Cheminformatics, QSAR and Machine Learning Applications for Novel Drug Development*, Elsevier, 2023, pp. 679–699. doi: 10.1016/B978-0-443-18638-7.00001-3.

[8] Plotly Technologies Inc., *Collaborative data science*. (2015). Montreal, QC. [Online]. Available: https://plot.ly

[9] W. Bittremieux, "spectrum_utils : A Python Package for Mass Spectrometry Data Processing and Visualization," *Anal. Chem.*, vol. 92, no. 1, pp. 659–661, Jan. 2020, doi: 10.1021/acs.analchem.9b04884.

[10] W. Bittremieux *et al.*, "Unified and Standardized Mass Spectrometry Data Processing in Python Using spectrum_utils," *J. Proteome Res.*, vol. 22, no. 2, pp. 625–631, Feb. 2023, doi: 10.1021/acs.jproteome.2c00632.

[11] OpenMS, *PyOpenMS-viz: Python Pandas-Based OpenMS Visualization Library*. [Online]. Available: https://github.com/OpenMS/pyopenms_viz

[12] OpenMS, *FLASHViewer*. [Online]. Available: https://github.com/t0mdavid-m/FLASHViewer

[13] P. Di Tommaso, M. Chatzou, E. W. Floden, P. P. Barja, E. Palumbo, and C. Notredame, "Nextflow enables reproducible computational workflows," *Nat. Biotechnol.*, vol. 35, no. 4, pp. 316–319, Apr. 2017, doi: 10.1038/nbt.3820.

[14] J. Köster and S. Rahmann, "Snakemake—a scalable bioinformatics workflow engine," *Bioinformatics*, vol. 28, no. 19, pp. 2520–2522, Oct. 2012, doi: 10.1093/bioinformatics/bts480.

[15] M. R. Berthold *et al.*, "KNIME - the Konstanz information miner: version 2.0 and beyond," *ACM SIGKDD Explor. Newsl.*, vol. 11, no. 1, pp. 26–31, Nov. 2009, doi: 10.1145/1656274.1656280.

[16] J. Goecks, A. Nekrutenko, J. Taylor, and T. Galaxy Team, "Galaxy: a comprehensive approach for supporting accessible, reproducible, and transparent computational research in the life sciences," *Genome Biol.*, vol. 11, no. 8, p. R86, 2010, doi: 10.1186/gb-2010-11-8-r86.

[17] L. M. Welp *et al.*, "Chemical crosslinking extends and complements UV crosslinking in analysis of RNA/DNA nucleic acid–protein interaction sites by mass spectrometry," Aug. 29, 2024. doi: 10.1101/2024.08.29.610268.

[18] OpenMS, *NASEWEIS*. [Online]. Available: https://github.com/poshul/nase-weis

[19] S. Wein and O. Kohlbacher, presented at the ASMS, 2024. [Online]. Available: www.openms.de/naseweis

[20] K. Dührkop, H. Shen, M. Meusel, J. Rousu, and S. Böcker, "Searching molecular structure databases with tandem mass spectra using CSI:FingerID," *Proc. Natl. Acad. Sci.*, vol. 112, no. 41, pp. 12580–12585, Oct. 2015, doi: 10.1073/pnas.1509788112.

[21] K. Dührkop *et al.*, "SIRIUS 4: a rapid tool for turning tandem mass spectra into metabolite structure information," *Nat. Methods*, vol. 16, no. 4, pp. 299–302, Apr. 2019, doi: 10.1038/s41592-019-0344-8.

[22] M. The, M. J. MacCoss, W. S. Noble, and L. Käll, "Fast and Accurate Protein False Discovery Rates on Large-Scale Proteomics Data Sets with Percolator 3.0," *J. Am. Soc. Mass Spectrom.*, vol. 27, no. 11, pp. 1719–1727, Nov. 2016, doi: 10.1007/s13361-016-1460-7.

[23] M. R. Lazear, "Sage: An Open-Source Tool for Fast Proteomics Searching and Quantification at Scale," *J. Proteome Res.*, vol. 22, no. 11, pp. 3652–3659, Nov. 2023, doi: 10.1021/acs.jproteome.3c00486.

[24] E. E. Kontou *et al.*, "UmetaFlow: an untargeted metabolomics workflow for high-throughput data processing and analysis," *J. Cheminformatics*, vol. 15, no. 1, p. 52, May 2023, doi: 10.1186/s13321-023-00724-w.

[25] L.-F. Nothias *et al.*, "Feature-based molecular networking in the GNPS analysis environment," *Nat. Methods*, vol. 17, no. 9, pp. 905–908, Sep. 2020, doi: 10.1038/s41592-020-0933-6.

[26] R. Schmid *et al.*, "Ion identity molecular networking for mass spectrometry-based metabolomics in the GNPS environment," *Nat. Commun.*, vol. 12, no. 1, p. 3832, Jun. 2021, doi: 10.1038/s41467-021-23953-9.

[27] A. K. Pakkir Shah *et al.*, "Statistical analysis of feature-based molecular networking results from non-targeted metabolomics data," *Nat. Protoc.*, Sep. 2024, doi: 10.1038/s41596-024-01046-3.

[28] S. Wein *et al.*, "A computational platform for high-throughput analysis of RNA sequences and modifications by mass spectrometry," *Nat. Commun.*, vol. 11, no. 1, p. 926, Feb. 2020, doi: 10.1038/s41467-020-14665-7.

# SUPPLEMENTARY

## Current OpenMS WebApps

### FLASHViewer

FLASHViewer is a comprehensive web application designed for visualizing and analyzing results from the FLASH* tool suite in top-down proteomics. It provides highly configurable, modular visualization components, enabling users to customize layouts and view multiple experiments simultaneously. Users can visualize various aspects, including raw and deconvolved spectrum plots, MS1 heatmaps, mass tables, 3D signal to noise plots, and dynamic sequence plots annotated with fragment ions. The application offers deep interactivity between components, such as clicking a row in a table to update related visualizations such as spectrum plots.

### NuXL-app

The NuXL-app is a web application version of the NuXL search engine, a specialized tool developed for the analysis of protein nucleic acid cross-linking mass spectrometry data. It allows for reliable, FDR-controlled assignment of protein–nucleic acid crosslinking sites from samples treated with UV light or chemical crosslinkers and offers user-friendly matched spectra visualization including ion annotations.

### UmetaFlow

This OpenMS WebApp offers the powerful UmetaFlow [24] pipeline for untargeted metabolomics in an accessible user interface. Raw data pre-processing converts raw data to a feature quantification table by feature detection, alignment, grouping, adduct annotation and optional re-quantification of missing values. Features can be annotated by in-house libraries based on MS1 m/z and retention time matching as well as MS2 fragment spectrum similarity. Features can be annotated with sum formula, compound name and chemical class by SIRIUS [20], [21]. For further investigation input files for GNPS feature-based molecular networking [25] and ion identity molecular networking [26] can be exported. Besides the untargeted pipeline, this app offers some additional features, such as an interface to explore raw data and metabolite identification and quantification via extracted ion chromatograms based on exact m/z values generated conveniently by an included m/z calculator. For downstream processing statistical analysis can be performed within the app or in the popular FBmn STATS GUIde [27].

## NASEWEIS

The NucleicAcidSearchEngine (also known as NASE) is a tool to do library searching and modification detection and localization on oligonucleotide mass spectrometry data [28]. It offers a wide variety of options to customize searching, which is ideal for experienced users but carries with it a learning curve. The NucleicAcidSearchEngine Web Execution In Streamlit (aka NASEWEIS), attempts to fill this gap by being a web-service version requiring no installation, geared for simple analysis of small experiments, and specifically only exposing the most important parameters to the user. NASEWEIS allows users to upload data from an MS experiment, a fasta file for (potentially modified) sequences to search, and set parameters regarding the resolution of the instrument. It then identifies candidates for Oligonucleotide spectral matches, and does FDR to return a table of Oligonucleotide hits, as well as an optional idXML output file for viewing the data with OpenMS' TOPPView.

## SagePTMScanner

Sage is a proteomics search engine that provides a variety of features such as retention time prediction, FDR-control, chimera searching and open searches among others [23]. SagePTMScanner provides an integration of Sage in OpenMS WebApps and aims to remove several hurdles towards using this software: it provides reasonable presets for searches, includes annotation and PTM discovery and produces immediate graphical output.

**Supplementary Figure 1. Running third party tools with OpenMS WebApps:** Third party tools can be used seamlessly with the WebApps framework. In this example MSConvert was integrated within the template. This mock workflow allows users to convert raw files to the mzML file format, select a single spectrum and m/z-range, and view the result as a graph. **A. File Upload:** Users can upload raw files. **B. Configure:** Users can select a single uploaded raw file, the spectrum ID and the m/z range. **C. Run:** MSConvert with the set parameters is run. **D. Results:** A simple plot of the resulting MSConvert output. **E. Sidebar:** The workflow can be navigated intuitively using the app's sidebar.